\documentclass[amsmath,aps,prc,twocolumn,groupedaddress, showkeys, showpacs]{revtex4}
\usepackage{graphicx}
\usepackage{epsfig}
\begin{document}

\title{Neutron specific heat in the crust of neutron stars from the nuclear band theory}

\author{N. Chamel}
\affiliation{Institut d'Astronomie et d'Astrophysique, Universit\'e
Libre de Bruxelles, CP226, Boulevard du Triomphe, 1050 Brussels, Belgium}

\author{J. Margueron}
\affiliation{Institut de Physique Nucl\'eaire, IN2P3-CNRS and Universit\'e Paris-Sud,
F-91406 Orsay Cedex, France}

\author{E. Khan}
\affiliation{Institut de Physique Nucl\'eaire, IN2P3-CNRS and Universit\'e Paris-Sud,
F-91406 Orsay Cedex, France}

\date{\today}

\begin{abstract}
The inner crust of neutron stars, formed of a crystal lattice of nuclear
clusters immersed in a sea of unbound neutrons, may be the unique example
of periodic nuclear systems. We have calculated the neutron specific heat
in the shallow part of the crust using the band theory of solids with Skyrme 
nucleon-nucleon interactions. We have also tested the validity of various 
approximations. We have found that the neutron specific heat is well described 
by that of a Fermi gas, while the motion of the unbound neutrons is strongly 
affected by the nuclear lattice. These apparently contradictory results are 
explained by the particular properties of the neutron Fermi surface.
\end{abstract}

\maketitle

Neutron stars are born in the gravitational core collapse of massive stars. 
With about one or twice the mass of the Sun compressed
inside a radius of only 10 kilometers or so, neutron stars are among the
most compact objects in the Universe~\cite{haen07}. The outer layers of 
the star, at densities below the neutron drip threshold 
$\rho_{\rm ND}\simeq 4\times 10^{11}$ g.cm$^{-3}$, are formed of a solid Coulomb
lattice of neutron rich nuclei coexisting with a degenerate gas of
relativistic electrons. The inner crust, at density above $\rho_{\rm ND}$ and 
below the crust-core transition density, which is at about half the
density $\rho_0 = 2.8 \times 10^{14}$ g.cm$^{-3}$ inside atomic nuclei, 
is permeated by a sea of unbound neutrons, which could be superfluid in 
some layers~\cite{lom99}. 

Many observed neutron star phenomena are intimately related to the physics 
of the crust. For instance, X-ray bursts in Low-Mass X-ray Binaries are 
associated with thermonuclear burning in the accreting
neutron star envelope~\cite{stro06}. In the subgroup of soft X-ray transients, accretion 
outbursts are followed by long period of quiescence during which the
accretion rate is essentially zero. In some cases, the period of accretion can
last long enough for the crust to be heated out of equilibrium with the
core. The thermal relaxation during the quiescent state has been
recently monitored for KS~1731$-$260 and for MXB~1659$-$29 after an
accretion episode of 12.5 and 2.5 years respectively~\cite{cac06}.  These
observations thus provide information on the thermal properties of neutron
star crusts~\cite{rut02, sht07}. The thermal relaxation of the crust could also be 
potentially observed in very young isolated neutron stars, $10-100$ years 
after their formation~\cite{lat94, gne01}. 

The diffusion of heat in the inner crust is mainly governed by the thermal
properties of the shallow layers, owing to their very low thermal
diffusivity~\cite{gne01}.  One of the key parameters is the neutron specific
heat. In cooling simulations of neutron stars, the neutron specific heat is generally 
approximated by that of uniform neutron matter. At low enough temperatures, 
neutrons are predicted to become superfluid but the exact value of the critical temperature $T_c$ still 
remains uncertain~\cite{lom99}. Moreover, the presence of nuclear clusters is likely to 
modify the critical temperature. For instance, in Ref.~\cite{gra08}, it has been shown that 
the pairing field in the dilute neutron gas completely vanishes just after the drip point.

The effects of superfluidity on the neutron specific heat are usually incorporated 
in the following way~\cite{lev94}:
\begin{equation}
\label{eq:cvpaired}
c_V^\mathrm{paired}(T)=\mathcal{R}(T/T_c) \,   c_V(T)
\end{equation}
where the renormalisation factor $\mathcal{R}(T/T_c)$ accounts 
for the effects of superfluidity and $c_V(T)$ stands for 
the specific heat calculated in the absence of superfluidity. 
In the above expression, the presence of the nuclear clusters is ignored. 
Their effects have been studied by solving the Hartree-Fock-Bogoliubov equations 
for the nucleons~\cite{piz02,monr07} and by including the
contribution of collective modes~\cite{khan05}. These calculations have
been carried out within the spherical Wigner-Seitz approximation. However
the validity of this approximation has been recently
discussed~\cite{cha07}. 
Pointing out the analogy between unbound neutrons in neutron star
crusts and conduction electrons in ordinary metals, 
the band theory of solids have been applied and adapted to the nuclear scale~\cite{car05,cha05}. 
It has shown for the first time the large effects of Bragg scattering on the motion of the unbound neutrons. 
One may expect that thermodynamical properties like the specific heat could also be modified.

In this paper, we investigate the effects of the solid crust on the neutron
specific heat, within the framework of the nuclear band theory
(Sec.~\ref{sec:bandtheory}). Given the present uncertainties in the 
pairing problem, even in the simpler case of pure neutron matter~\cite{lom99}, we 
will focus on the normal part $c_V(T)$ only. 
Results in the shallow layers are discussed in Sec.~\ref{sec:results} and are compared to 
several approximations used in the literature.

\section{A microscopic model of the neutron star crust}
\label{sec:bandtheory}

Following the standard assumptions, we consider a body centered cubic crystal with only
one nuclear cluster per lattice site~\cite{haen07}. We have taken 
the composition $(N,Z)$ of the clusters calculated by Negele and Vautherin~\cite{ne73}. 
But the nucleon distributions have been recalculated at each temperature $T$ by 
solving the finite temperature Hartree-Fock equations with the Skyrme SLy4 effective
nucleon-nucleon interaction~\cite{chabanat98}. Typical temperatures of interest 
for cooling isolated neutron stars and for accreting neutron stars lie in the range 
between 10$^7$ to few 10$^9$~K~\cite{gne01}.

Fully self-consistent calculations within the band theory are
computationally very expensive. We have thus solved the Hartree-Fock
equations in two steps~\cite{cha07}.  First, we have determined
self-consistently the nucleon distributions and mean fields in the
Wigner-Seitz approximation with mixed Neumann-Dirichlet boundary conditions. We have 
removed the spurious fluctuations of the neutron density by averaging in the
interstitial region~\cite{ne73}. In a second stage, we have solved the
Schr\"odinger equation with Bloch boundary conditions using these
self-consistent mean fields. We have fixed the lattice spacing so that the
volume of the Wigner-Seitz polyhedron is equal to the volume of the
spherical Wigner-Seitz cell of radius $R_{\rm cell}$.  We have employed the
Linearized Augmented Plane Wave method described in details in
Ref.~\cite{cha05}. 
As shown in Ref.~\cite{cha07}, the one-body spin-orbit potential 
has a negligible effect on the neutron level ordering 
and was therefore neglected.

\begin{table}[b]
\centering
\setlength{\tabcolsep}{.05in}
\renewcommand{\arraystretch}{1.8}
\caption{For the three shallow layers considered in this paper, are given:
the total density ($\rho$), the number of protons and neutrons ($Z$, $N$ respectively),
the nuclear mass number $A$ (at $T=0$), the radius of the cell ($R_\mathrm{cell}$) and 
the neutron gas density at $T$=0
($\rho_n^G$).}
\label{tab.struc}
\vspace{.5cm}
\begin{tabular}{cccccc}
\hline
\hline
$\rho$ [g.cm$^{-3}$] & $Z$ & $N$ & $A$ & $R_{\rm cell}$ [fm] & $\rho_n^G$ [fm$^{-3}$] \\
\hline
$6.69 \times 10^{11}$ & 40 & 160 & 133 & 49.24 & 1.3 10$^{-4}$ \\
$1.00 \times 10^{12}$ & 40 & 210 & 141 & 46.33 & 2.6 10$^{-4}$ \\
$1.47 \times 10^{12}$ & 40 & 280 & 140 & 44.30 & 4.9 10$^{-4}$ \\
\hline
\hline
\end{tabular}
\end{table}

The specific heat (per unit volume) is defined as
\begin{equation}
\label{eq.cv}
c_V(T)=\frac{\partial U}{\partial T} \biggr\vert_V=
T \frac{\partial S}{\partial T}\biggr\vert_V \, ,
\end{equation}
(taking the Boltzmann's constant $k_{\rm B}=1$)
where $U$ is the total internal energy density, $S$ the entropy density and the 
partial derivatives are evaluated at constant volume $V$. 
We have found that the latter expression is numerically more convenient. 
The entropy density of the unbound neutrons is given by 
\begin{equation}
S=-\sum_\alpha\hspace{-0.1cm}\int\hspace{-0.1cm}\frac{{\rm d}^3\pmb{k}}{(2\pi^3)} 
\biggl[ f_{\alpha\pmb{k}} \ln  f_{\alpha\pmb{k}}  + 
(1- f_{\alpha\pmb{k}} )\ln (1-f_{\alpha\pmb{k}} )\biggr]
\label{eq:s}
\end{equation}
where $\alpha$ is the band index, $\pmb{k}$ is the Bloch wave vector and $f_{\alpha\pmb{k}}$ 
is the Fermi-Dirac distribution. 
The latter depends on the temperature and the neutron chemical potential $\mu_n$ which is 
determined from the total density $\rho^G_n$ of unbound neutrons by
\begin{equation}
\rho_n^G=\sum_\alpha \int \frac{{\rm d}^3\pmb{k}}{(2\pi^3)} f_{\alpha\pmb{k}} \, .
\end{equation}
Integrations have been carried out using the special point method 
(see \cite{cha05}).
Note that the number of unbound neutrons per cluster is determined by the shape of the 
mean fields, which vary with temperature. Consequently $\rho^G_n$ can also depend 
on the temperature. In particular, with increasing temperature the most loosely bound 
neutrons may be excited into the delocalized states. 

\section{Results and discussions}
\label{sec:results}

We have considered three different layers in the shallow region of the inner
crust. Their properties are summarized in Table~\ref{tab.struc}. 
The neutron chemical potential $\mu_n(T)$ and the entropy density $S(T)$, 
have been calculated for a set of temperatures between $T=10$ keV and $T=100$
keV.  We have
then evaluated the specific heat according to Eq.~(\ref{eq.cv}), 
by interpolating the entropy density $S(T)$
with cubic splines and numerically differentiating. 

For the range of temperatures considered here,
$T\ll \varepsilon_\mathrm{F}$, where 
$\varepsilon_\mathrm{F}$ is the Fermi energy. 
The specific heat is then approximately given by~(ref.~\cite{am76}, p47)
\begin{equation}
\label{eq.cv-app}
c_V(T)\simeq \frac{\pi^2}{3} g(\varepsilon_\mathrm{F}) T \, ,
\end{equation} 
where $g(\varepsilon_\mathrm{F})$ is the density of states at the Fermi level. 
This expression is valid provided i) the density of unbound neutrons $\rho^G_n$
and $\varepsilon_\mathrm{F}$ are independent of the temperature, 
ii) the density of states $g(\varepsilon)$ is sufficiently smooth.
Pertaining to the first point, we have found numerically that $\rho_n^G$ varies very little 
so that the neutron evaporation/condensation phenomenon mentionned in 
Sect.~\ref{sec:bandtheory} can be 
ignored. In a previous paper, we have shown that the density of states in the shallow 
layers of the crust is far from being smooth at a scale of few keV (see Fig.~6 of 
Ref.~\cite{cha07}). However, since we consider thermal energies of order 
$10$ keV or more, the variations of the density of states with the energy 
are expected to be much smaller at this scale as noticed in 
Ref.~\cite{cha07}. As can be seen in Fig.~\ref{fig.cv}, the neutron specific heat varies 
almost linearly with $T$ as expected from Eq.~(\ref{eq.cv-app}). 

In order to understand more qualitatively our results, we have performed a 
comparison to several approximations that have been used in the literature.

\subsection{Validity of various approximations}
\label{sect:approx}

\begin{figure}
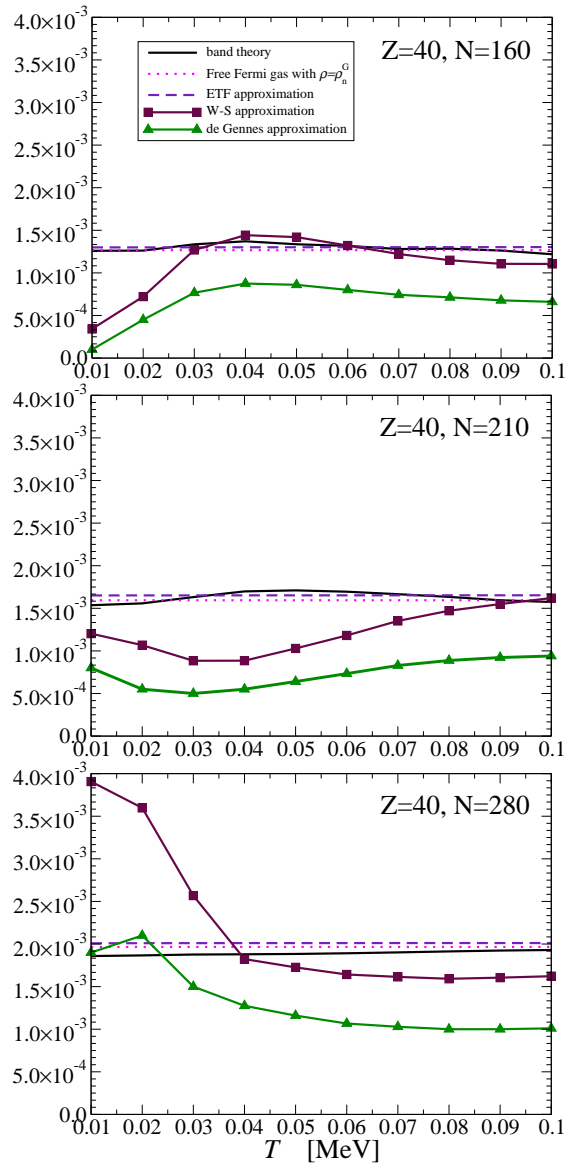

\begin{center}
\epsfig{file=cv1d.eps,scale=0.3}
\epsfig{file=cv2d.eps,scale=0.3}
\epsfig{file=cv3d.eps,scale=0.3}
\end{center}
\caption{ (color online) Neutron specific heat divided by the temperature in the three shallow layers
calculated with the band theory (solid thick line) and with different approximations: 
free Fermi gas $c_V^\mathrm{FG}$~(\ref{eq.cvfermi}) (dotted line), 
Extended-Thomas-Fermi $c_V^\mathrm{TF}$~(\ref{eq.cv-tf}) (dashed line),
Wigner-Seitz $c_V^\mathrm{WS}$~(\ref{eq.cv-ws}) (solid line with filled squares) and
de Gennes $c_V^\mathrm{DG}$~(\ref{eq.cv-dg}) (solid line with
filled triangle). 
}
\label{fig.cv}
\end{figure}

The simplest approximation to the neutron specific heat 
is to neglect the presence of the nuclear clusters. 
The specific heat of a Fermi gas is given by
\begin{equation}
\label{eq.cvfermi}
c_V^\mathrm{FG}(T)= \left(\frac{\pi}{3}\right)^{2/3}
\frac{m_n^\oplus T}{\hbar^2} \left(\rho_n^G\right)^{1/3} \; ,
\end{equation}
where $m_n^\oplus$ is the Skyrme effective mass evaluated at $T=0$ for 
the density of the neutron gas $\rho_n^G$. Note that for the layers we considered, 
$m_n^\oplus\simeq m_n$. 
As shown in Fig.~\ref{fig.cv}, the specific heat given by~(\ref{eq.cvfermi}) 
is very close to the exact result from the band theory. 

This striking result can be understood using the semi-classical Extended-Thomas-Fermi 
expansion method~\cite{onsi08}. At lowest order in $\hbar^2$, the neutron specific heat is given by 
\begin{equation}
\label{eq.cv-tf}
c_V^\mathrm{TF}(T)= \left(\frac{\pi}{3}\right)^{2/3} \frac{T}{\hbar^2} 
\int \frac{{\rm d}^3 \pmb{r}}{V_\mathrm{cell}} m^{\oplus}_n(\pmb{r}) \rho_n(\pmb{r})^{1/3} \, .
\end{equation} 
We have found that higher order corrections in $\hbar^2$ are negligible.
The specific heat~(\ref{eq.cv-tf}) shown in 
Fig.~\ref{fig.cv} for the three selected layers, is very close to the 
results of band theory. The small effects of the nuclear lattice can now be easily 
explained by the small volume fraction occupied by the clusters (typically $\sim 10^{-2}-10^{-3}$). 
Indeed in the limit of uniform neutron matter, the Thomas-Fermi specific heat~(\ref{eq.cv-tf})
reduces to Eq.~(\ref{eq.cvfermi}). A more refined explanation 
within the full quantum mechanical framework will be given in Sect.~\ref{sec:fermigas}.

We have also computed the specific heat with the single particle 
energies $\varepsilon_\alpha$ obtained in the Wigner-Seitz approximation.
Varying the temperature, we have found a small redistribution of the
neutron energies $\varepsilon_\alpha$ close to the Fermi energy.
Due to the very large number of neutron states, a small redistribution 
of energies can have a large impact on the total energy 
and on the entropy. 
However, it has been found numerically that the entropy being a 
smooth function of the occupied and unoccupied states, is much less
affected by this redistribution than the total energy.
We thus have calculated the specific heat using 
\begin{eqnarray}
\label{eq.cv-ws}
c_V^\mathrm{WS}&=& T\frac{\partial S^\mathrm{WS}}{\partial T}
\biggr\vert_V\; .
\end{eqnarray} 
The entropy density is given by
\begin{eqnarray}
\label{eq.s-ws}
S^\mathrm{WS}= -\frac{1}{V_\mathrm{cell}}\sum_{\alpha} g_\alpha \biggl[f_\alpha \ln f_\alpha
+(1-f_\alpha) \ln (1-f_\alpha)\biggr] 
\end{eqnarray} 
where $g_\alpha$ is the degeneracy of the state $\alpha$.
The specific heat~(\ref{eq.cv-ws}) is shown in Fig.~\ref{fig.cv}.
Despite non-linear fluctuations due to the redistribution of neutron states,
the results are fairly close to those of the band theory. 
The best agreement is obtained for the layer with the lowest density, 
as expected from the domain of validity of the Wigner-Seitz method~\cite{cha07},

We have checked the validity of another approximation, suggested by 
de Gennes~\cite{degen66} and applied in Ref.~\cite{monr07}, consisting 
in differentiating the entropy~(\ref{eq.s-ws}) with respect
to the temperature but assuming that the energies 
$\varepsilon_\alpha$ and the chemical potential $\mu_n$ (as well as the pairing gaps in superfluid systems)
are independent of $T$. The specific heat is then given by
\begin{equation}
\label{eq.cv-dg}
c_V^\mathrm{DG}(T)=\frac{1}{V_\mathrm{cell}} \sum_\alpha  g_\alpha
f_\alpha (1- f_\alpha) \biggl(\frac{\varepsilon_\alpha-\mu_n}{T}\biggr)^2 \, .
\end{equation}
As can be seen in Fig.~\ref{fig.cv}, this approximation is valid at very 
low temperature where the energies and the chemical potential could be 
taken independent of the temperature but it becomes less and less
reliable with increasing temperature.

From the comparison of various approximations to the neutron specific heat, 
we have found that the Fermi gas model reproduces very well
the results of the band theory, while more sophisticated approaches like the 
Wigner-Seitz or the de Gennes approximations are less accurate. 
Does it mean that the unbound neutrons are really ``free''?

\subsection{Are unbound neutrons really ``free''?}
\label{sec:fermigas}

The present results on the specific heat together with Eq.~(\ref{eq.cv-app}) 
indicate that the average density of states are close to that of a Fermi gas 
(as already noticed in a previous paper, see Fig.~6 and 9 of Ref.~\cite{cha07}). 
This conclusion is at first sight surprising 
since the mean field potential is very deep inside the clusters (about -70 MeV). On general 
grounds, one expects the effects of the nuclear lattice to be negligible 
whenever the Fermi wavelength of the unbound neutrons 
$\lambda_\mathrm{F}=2\pi/k_\mathrm{F}$ with $k_\mathrm{F}=(3\pi^2\rho_n^G)^{1/3}$, 
is much larger than the lattice spacing.  However this condition is not 
satisfied for the layers we considered. Using the densities $\rho_n^G$ given 
in Table~\ref{tab.struc}, we find $\lambda_\mathrm{F}\simeq 40.09$, $31.82$ 
and $25.76$ fm for $R_\mathrm{cell}=49.24$, $46.33$ and $44.3$ fm, respectively. 
One therefore expects the unbound neutrons to be strongly scattered by the clusters. 

If the unbound neutrons were really free as suggested by the results on the 
specific heat, this would imply that their motion is unaffected by the lattice. 
It is well-known in solid state physics that the interactions between the conduction
electrons and the ionic lattice can be taken into account 
through renormalizating the mass of the electrons. 
The motion of an electron in a solid, with a wave vector 
$\pmb{k}$ in a band $\alpha$, can thus be characterized 
with a dynamical effective mass tensor defined by (ref.~\cite{am76}, p228) 
\begin{equation}
\frac{1}{m_e^\star}\biggr\vert_{ij} = \frac{1}{\hbar^2} 
\frac{\partial^2\varepsilon^{(e)}_{\alpha\pmb{k}}}{\partial k_i \partial k_j}
\end{equation}
where $k_i$, $i=1,2,3$ are the Cartesian components of $\pmb{k}$ and
$\varepsilon^{(e)}_{\alpha\pmb{k}}$ the electron
energy. This concept has been extended to low-energy thermal neutrons
propagating in crystals and the corresponding effective mass has been
experimentally measured in silicon~\cite{Zeilinger86}. In the
context of neutron star crust, since the number of unbound neutrons per 
lattice site can be very large, it is more appropriate to introduce an 
average effective mass $m_n^\star$ defined by~\cite{car05}
\begin{equation}
\label{eq.dyneffmass1}
m_n^\star = \rho_n^G/{\cal K} \, , \hskip0.5cm {\cal K}=\frac{1}{3}\sum_{\alpha, i} \int_{\rm F} 
\frac{{\rm d}^3\pmb{k}}{(2\pi)^3}\, \frac{1}{\hbar^2} 
\frac{\partial^2\varepsilon_{\alpha\pmb{k}}}{\partial k_i\partial k_i} \, ,
\end{equation}
where the integral is taken over all occupied states. This
dynamical effective mass $m_n^\star$ can be equivalently expressed as
\begin{equation}
\label{eq.dyneffmass2}
\frac{1}{m_n^\star} = \frac{1}{\rho_n^G}\sum_i\frac{\partial^2 U}{\partial p_{ni} \partial p_{n i}} \, ,
\end{equation}
where $U$ is the energy density of the moving neutrons in the crust frame 
and $\pmb{p_n}$ is the average neutron momentum. This effective mass has 
implications for neutron star dynamics. For instance it has been shown that 
a large enough effective mass can trigger a Kelvin-Helmholtz instability 
which might explain the origin of pulsar glitches~\cite{and03}. 

The dynamical effective mass $m_n^\star$ has been calculated for the 
considered layers, using the same numerical method as in Ref.~\cite{cha05}. 
Results are shown in Table~\ref{tab.effmass}. 
If the unbound neutrons were free we would have found $m^\star_n/m_n=1$. 
However, the dynamical effective mass is much larger than the bare neutron mass 
$m^\star_n/m_n>1$ indicating that the interactions between the neutrons and the clusters
are very strong. In the following, we will explain the apparent contradiction with the
results obtained previously for the specific heat. 

At the beginning of Sect.~\ref{sec:results}, 
we have shown that the specific heat varies almost linearly with 
the temperature (see Fig.~\ref{fig.cv}), as expected from Eq.~(\ref{eq.cv-app}).
The density of states $g(\varepsilon_\mathrm{F})$ appearing in Eq.~(\ref{eq.cv-app})
can be expressed as an integral over the Fermi surface $\mathcal{S}_\mathrm{F}$,
\begin{equation}
\label{eq.level-density}
g(\varepsilon_{\rm F})=\frac{1}{(2\pi)^3\hbar} \sum_\alpha \oint_{\mathcal{S}_\mathrm{F}} 
\frac{d {\cal S}}{|\pmb{v}_{\alpha\pmb{k}}|} \, .
\end{equation}
where $\pmb{v}_{\alpha\pmb{k}}=\hbar^{-1} \nabla_{\pmb{k}}\varepsilon_{\alpha\pmb{k}}$ is 
the group velocity of the unbound neutrons. 
The presence of nuclear clusters leads to the existence of resonances. As a result, 
the energy bands of unbound states may be locally flat in k-space thus distorting the 
Fermi surface~\cite{cha07}. 
The deformation can be estimated by the ratio $\xi_\mathrm{F}$ of the area of 
the deformed surface to that of the unperturbed Fermi sphere. Therefore, we have 
${\cal S}_{\rm F} = \xi_\mathrm{F} 4\pi k_\mathrm{F}^2$, where by definition $\xi_\mathrm{F}=1$ in the absence of 
clusters. Numerical calculations show that the nuclear lattice reduces the 
area of the Fermi surface by more than a factor of 2 in the shallowest layer 
(see Table~\ref{tab.effmass}). 
However since the specific heat is driven by that of a Fermi gas, we can infer from 
Eq.~(\ref{eq.level-density}) that in average, the group velocity is changed by the same 
factor $\xi_\mathrm{F}$ compared to that of the Fermi gas. 

\begin{table}[b]
\centering
\setlength{\tabcolsep}{.05in}
\renewcommand{\arraystretch}{1.8}
\caption{For the three shallow layers considered in this paper, are given:
the total density ($\rho$), the dynamical effective mass ($m^\star_n/m_n$) and the reduction factor of
the Fermi surface area ($\xi_F$). }
\label{tab.effmass}
\vspace{.5cm}
\begin{tabular}{cccc}
\hline
\hline
$\rho$ [g.cm$^{-3}$] & $m_n^\star/m_n$ & $\xi_F$ & $\xi_F^2 m_n^\star/m_n$ \\
\hline
$6.69 \times 10^{11}$ & 4.0 & 0.44 & 0.77 \\
$1.00 \times 10^{12}$ & 3.6 & 0.49 & 0.86\\
$1.47 \times 10^{12}$ & 3.2 & 0.52 & 0.87\\
\hline
\hline
\end{tabular}
\end{table}

Likewise the coefficient $\cal K$ (\ref{eq.dyneffmass1}) can also be written as an 
integral over the Fermi surface
\begin{equation}
{\cal K}=\frac{1}{3 (2\pi)^3\hbar}\sum_\alpha\oint_\mathcal{S_\mathrm{F}} 
|\pmb{v}_{\alpha\pmb{k}}| {\rm d}{\cal S}\, .
\end{equation}
From the previous discussion, it follows that this coefficient is therefore 
approximately changed by a factor $\xi_\mathrm{F}^2$. Eq.~(\ref{eq.dyneffmass1}) then 
implies that $m_n^\star \approx m_n/\xi_\mathrm{F}^2$. 
Inspecting Table~\ref{tab.struc} shows that the predicted relation between 
$m_n^\star$ and $\xi_\mathrm{F}$ is only roughly satisfied 
because $v_{\alpha\pmb{k}}$ varies on the Fermi surface. 
Nevertheless, this analysis in terms of the topology of the Fermi surface 
explains why the presence of the nuclear clusters has such a strong impact 
on the motion of the unbound neutrons but not on their specific heat. 

\section{Conclusions}

Modelling the recently observed thermal relaxation of neutron star crusts
requires the knowledge of their thermal properties. In this paper, 
we have computed the specific heat of the neutron ocean permeating the inner 
crust at densities below $1.5\times 10^{12}$ g.cm$^{-3}$, 
by applying the band theory of solids with the Skyrme SLy4 nucleon-nucleon 
interaction. We have compared the results obtained using different approximations. 
We have found that for temperatures $T=10^7-10^9$K relevant to 
neutron stars, the neutron specific heat is essentially given by that of a Fermi gas.
The Thomas-Fermi expression~(\ref{eq.cv-tf}) yields nearly undistinguishable results. The specific
heat calculated in the Wigner-Seitz approximation agrees reasonably well, while the 
de Gennes approximation~(\ref{eq.cv-dg}) leads to specific heats a factor of 2-3 smaller 
than those obtained in the band theory. 

The results on the specific heat might suggest that the unbound neutrons are not affected 
by the presence of the nuclear clusters. 
It is however not true since the dynamical effective mass of the same unbound neutrons
is very different from the bare one
unvealing strong interactions with the periodic lattice. 
This apparent paradox can be explained by the 
fact that the neutron Fermi surface area and the neutron group velocity are both reduced by the same 
factor in the presence of clusters, leading to a nearly unchanged density of states
and specific heat.

\begin{acknowledgments}
N. C. gratefully acknowledges financial support from a Marie Curie Intra-European grant 
(contract number MEIF-CT-2005-024660) and from FNRS (Belgium). 
\end{acknowledgments}

\end{document}